\documentclass[12pt,a4paper]{iopart}

\usepackage{graphicx,subfig}

\newcommand{\bmx}{Ba$M_2X_2$}

\newcommand{\bcu}{BaCu$_2$S$_2$}
\newcommand{\bce}{BaCu$_2$Se$_2$}
\newcommand{\bau}{BaAg$_2$S$_2$}
\newcommand{\bae}{BaAg$_2$Se$_2$}

\begin{document}

\title{First-principles study of band alignments in the $p$-type hosts \bmx\/ ($M$ = Cu, Ag; $X$ = S, Se)} 

\author{Aditi Krishnapriyan, Phillip T. Barton, Maosheng Miao, and Ram Seshadri} \eads{seshadri@mrl.ucsb.edu}
\address{Materials Research Laboratory\\
University of California, Santa Barbara, CA, 93106, USA}

\date{\today}

\begin{abstract}
The electronic structures of four semiconductor compounds \bcu\/, \bce\/, \bau\/, and \bae\/ are studied by density functional theory using both semi-local and hybrid functionals. The ionization energies and electron affinities were determined by aligning the electronic states with the vacuum level by calculating the electrostatic profile within a supercell slab model. The ionization energy and electron affinity of the compounds were calculated using the Heyd-Scuseria-Ernzerhof (HSE) functionals and range from 4.5 to 5.4\,eV and 3.1 to 3.4\,eV, respectively. The replacement of Cu by Ag slightly increases the ionization energy and electron affinity, while the replacement of S by Se decreases the ionization energy but slightly increases the electron affinity. Overall, the low ionization energies and small electron affinities suggest that these compounds possess good $p$-type doping propensities. The band gaps are somewhat small to be ideal candidates for transparent semiconducting behavior. The replacement of Cu with Ag in the barium sulfide compounds can increase the band gap from 1.62\,eV to 2.01\,eV.

\end{abstract}
\maketitle

\section{Introduction}

Transparent conducting oxides (TCO) are important materials in the development of energy-related devices such as solar cells and light emitting diodes. One of the major issues with these TCOs is the lack of efficient $p$-type TCO materials, which prevents the creation of transparent electronic devices.\cite{Thomas_Nature97,Zunger_APL03} Traditional TCO materials such as In$_{2}$O$_{3}$ and SnO$_{2}$ are readily doped $n$-type, but have yet to be successfully doped $p$-type.\cite{Ingram_JEC04} This large doping asymmetry originates from a common electronic structure feature of these TCO materials, namely that their highest energy valence states are mainly O-$p$ and their lowest conduction states are mainly cation $s$. These materials therefore possess large ionization energies and high electron affinities, making them strongly resistive to $p$-type doping. 

In order to improve the $p$-type doping propensity, the nature of the filled valence bands needs to be changed. One approach is to introduce filled $M$-$d$ levels which are usually higher in energy than the O-$p$ level. This occurs in Cu$_2$O, for example, which is $p$-type due to intrinsic defects that produce hole states. While these holes are relatively mobile, the band gap of Cu$_{2}$O is only 2.1\,eV,\cite{Raebiger_PRB07} which is too low for a TCO. Cu-based delafossites, such as CuAlO$_2$\cite{Kawazoe_Nature97}, have larger gaps due to the 2D nature of their crystal structure, but their application as TCOs is impaired by their low hole mobility which is also a result of the layered structure.\cite{Yanagi_JAP00, Ingram_PRB01} This mobility drawback makes another approach attractive: replacing O with another chalcogen element such as S or Se. While this could result in more disperse bands, it is possible that this would also come at the price of a smaller band gap. The $p$ states of these elements are usually higher, making their compounds more prone to being $p$-type doped. Recently, some effort has been made to investigate the potential of using Cu-containing chalcogenides, such as LaOCuS\cite{Ueda_APL00} and BaCu$_{2}$S$_{2}$\cite{Park_APL02}, as $p$-type transparent semiconductors. 

The electronic structures of these chalcogenide compounds have not been thoroughly studied by first principles methods. BaCu$_{2}$S$_{2}$ was recently examined using semi-local functionals and found to have a calculated band gap of 0.50\,eV,\cite{Liu_PBCM10} however these functionals typically underestimate the band gap and are not well-suited to assess the materials potential as a transparent semiconductor. On the other hand, the experimental measurement of 2.3\,eV for BaCu$_{2}$S$_{2}$ might be larger than its actual gap because of the exceedingly low density of states at the conduction band edge and the small optical oscillator strength.\cite{Park_APL02} Furthermore, band alignments of semiconductor materials can provide guidelines in understanding their doping propensity. Generally, materials with smaller ionization energy are prone to $p$-type doing, whereas materials with larger electron affinity are prone to $n$-type doping. 

%To address this problem of low hole mobility, an effort has been made to look at chalcogenides (S, Se), where the Cu-S interactions in compounds such as LaOCuS have been shown to have higher covalency than the Cu-O interactions.} Sulfides are of interest because the larger 3$p$ orbitals may lead to a more disperse valence band and higher hole mobility, though this would likely come at the price of a smaller band gap.

%The characteristic properties of semiconducting materials that are important for obtaining $p$-type transparent conducting include the ionization energy and the effective masses at the band edge. A smaller ionization energy indicates the propensity of p-type doping whereas the effective masses strongly relate to the hole mobility. A higher covalency should lead to smaller effective masses and therefore higher mobilities, which should give improved conductivity. 

In this contribution, the electronic structures of \bmx\/ ($M$ = Cu, Ag; $X$ = S, Se) compounds have been calculated using the density functional theory (DFT) method with both semi-local and hybrid functionals. The Ag compounds are not known experimentally. The bands of these compounds are aligned with the vacuum level by calculating the electrostatic potential profile of a slab/vacuum supercell. The results show that the compounds have small ionization energies, indicating a potential ability to be $p$-type doped, but that their electronic band gaps are smaller than desired for TCOs. The substitution of Ag for Cu opens the band gaps slightly while retaining nearly the same ionization energy.

% Calculate effective masses. Can we quantify the M-d and X-p mixing at the VBM?

\section{Computational methods}

The electronic structure calculations and geometry optimizations are based on the density functional theory (DFT) method as implemented in the Vienna ab-Initio Simulation Project (VASP) code.\cite{VASP} The projector augmented wave (PAW) potentials are used to describe the ionic potentials of the elements. In order to examine the effect of the different choice of functional, both semi-local Perdew-Burke-Ernzerhof (PBE) generalized gradient approximated (GGA) functionals and the Hyed-Scuseria-Ernzerhof (HSE) hybrid functional were employed.\cite{PBE,HSE} An 8$\times$8$\times$4 Monkhorst-Pack $k$-mesh was used for all of the PBE calculations, whereas a 6$\times$6$\times$2 $k$-mesh was used for HSE calculations. A cutoff energy of 550\,eV was found to be sufficient to obtain accurate forces and stresses for the geometry optimizations.  

\begin{figure}
\centering
\includegraphics[width=1.7in]{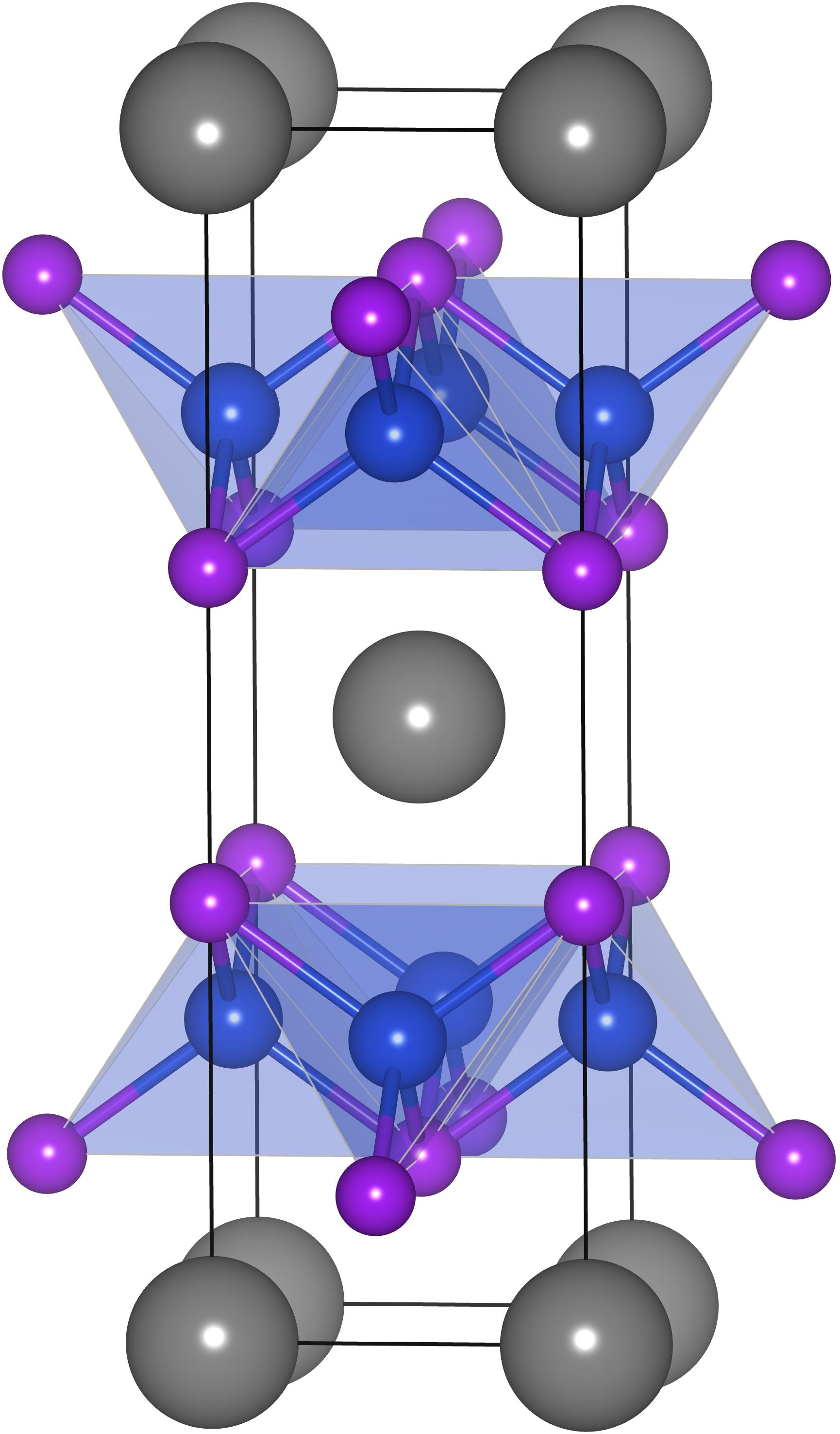}
\caption{Schematic ThCr$_2$Si$_2$-type crystal structure of \bmx, tetragonal space group $I$4/$mmm$. Ba, $M$, and $X$ are colored grey, blue, and purple and positioned at (0, 0, 0), (0, 0.5, 0.25), and (0, 0, $x \approx$ 0.36) respectively.}
\label{fig:structure}
\end{figure}

\begin{figure}
\centering
\includegraphics[width=3.5in]{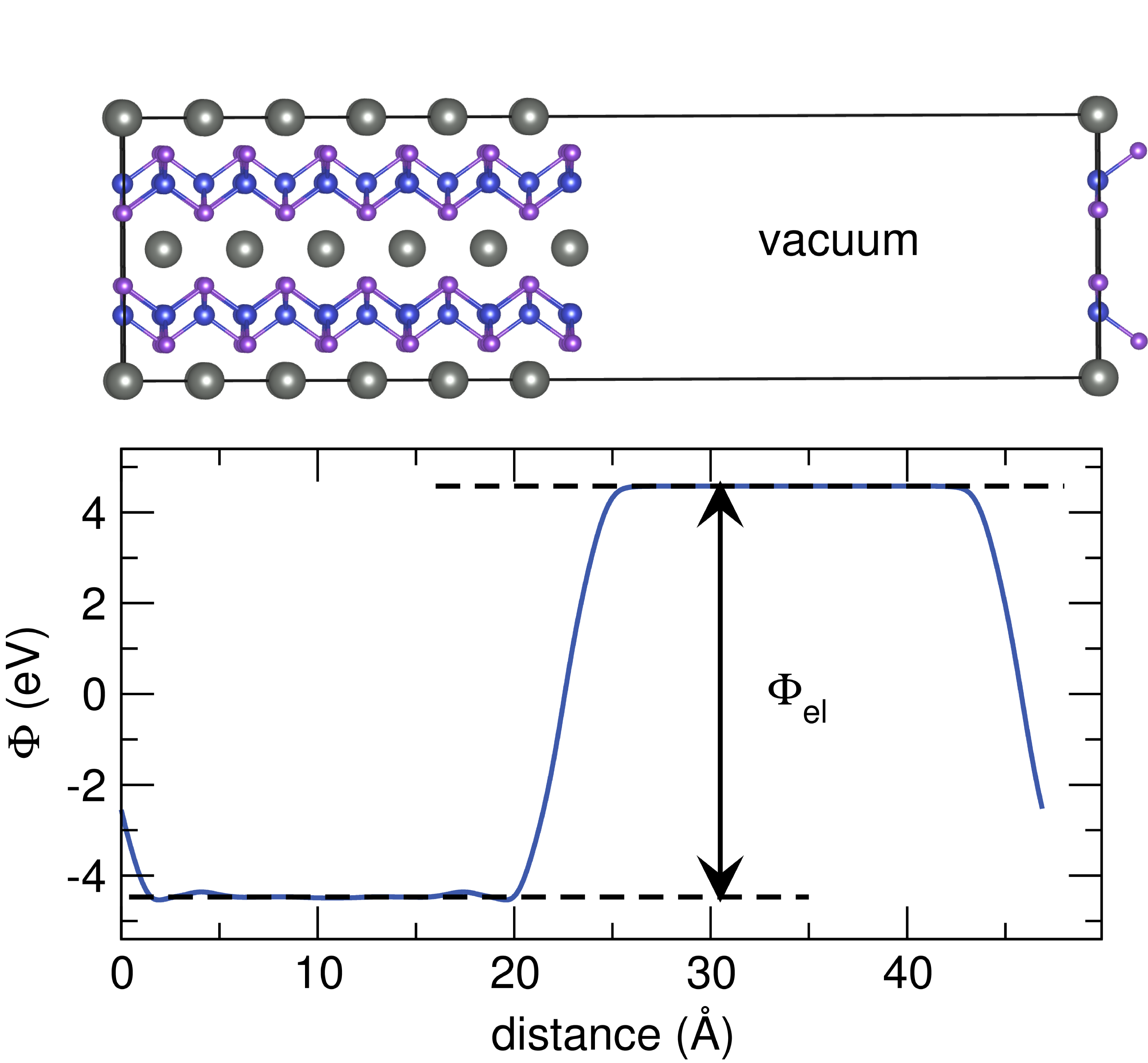}
\caption{Schematic slab model (top), where atoms on the far right represent the periodic boundary conditions, and the corresponding calculated electrostatic potential (bottom). The potential difference between the bulk and vacuum determines the absolute position of the band edges, and these band edge positions are directly related to the ionization energy and electron affinity.}
\label{fig:slab}
\end{figure}

Slab model calculations were performed to find the alignment of the valence band maximum (VBM) and conduction band minimum (CBM) relative to the vacuum level.\cite{Kurzman_JPCM11,Moses_JCP11} A supercell was constructed and stacked along a non-polar direction ($y$ as shown in Fig.~\ref{fig:slab}) using the optimized atomic structure obtained by the PBE functional. The supercell consisted of a bulk region of six unit cells and a vacuum region of the same length. The electrostatic potential of the bulk region was then aligned with the vacuum region center to determine the alignment of the VBM and CBM relative to the vacuum. The ionization energy $E_{\rm{I}}$ and the electron affinity $E_{\rm{A}}$ are calculated by:
\begin{eqnarray}
E_{\rm{I}}=\Phi_{\rm{el}} - E_{\rm{VBM}} \\ \nonumber
E_{\rm{A}}=\Phi_{\rm{el}} - E_{\rm{CBM}}
\end{eqnarray}
in which $E_{\rm{VBM}}$ and $E_{\rm{CBM}}$ are the absolute values of the VBM and CBM energies obtained from bulk calculations, and $\Phi_{\rm{el}}$ is the electrostatic potential difference between the bulk center and the vacuum as calculated from the slab model. 

\section{Results and Discussion} 

\subsection{Crystal Structure}

%Liu, GB et al used DFT to find a band gap of .97 eV for alpha BaCu2S2, .50 eV for beta-BaCu2S2
The chalcogenide \bmx\/ compounds are composed of divalent cations (Ba$^{2+}$), monovalent transition metal cations (Cu$^{+}$ or Ag$^{+}$), and divalent anions (S$^{2-}$ or Se$^{2-}$). They crystallize in the ThCr$_2$Si$_2$ structure type ($I$4/$mmm$ space group), consisting of layers of edge-sharing $MX_4$ tetrahedra, separated by Ba$^{2+}$ cations. The Ag compounds have not been synthesized experimentally, but are investigated here theoretically. BaAg$_2$S$_2$ is known to form in the La$_2$O$_3$ structure type.\cite{Bronger_ZAAC97} The total energy calculations find that BaAg$_2$S$_2$ is more stable in its own hypothetical structure than in the ThCr$_2$Si$_2$ structure by 0.17\,eV/cell. Table~\ref{geometry} lists the experimental and calculated values of the lattice constants, the $z$ position of $X$, and the $M$-$X$ bond lengths. Both PBE and HSE geometry parameters are in good agreement with the experimental values. Replacements of both Cu and S with larger ions lead to increasing lattice parameters as well as increasing $M$-$X$ distances. The replacement of Cu by Ag has a more significant effect than of S by Se, which is consistent with a larger radii difference for Cu$^{+}$ (0.60 \AA) and Ag$^{+}$ (1.00 \AA) ions than for S$^{2-}$ (1.84 \AA) and Se$^{2-}$ (1.98 \AA) ions.\cite{Shannon_ACB69}

\begin{table*}[t]\footnotesize
\begin{center}
\caption{The calculated and experimental lattice constants ($a$ = $b$), $c/a$ ratio, $z$ position of $X$, and $M$-$X$ bond lengths for \bmx\/ ($M$ = Cu, Ag; $X$ = S, Se) compounds.} 
\label{geometry}
\vspace{4pt}
\begin{tabular}{p{0.8cm}p{0.1cm}p{0.7cm}p{0.7cm}p{0.7cm}p{0.1cm}p{0.7cm}p{0.7cm}p{0.7cm}p{0.1cm}p{0.7cm}p{0.7cm}p{0.7cm}p{0.1cm}p{0.7cm}p{0.7cm}p{0.7cm}p{0.1cm}p{0.7cm}p{0.7cm}p{0.7cm}}
\hline
~&~& & $a$~=~$b$~(\AA)& & & & $c/a$ & & & & $z$ & & & & \mbox{$M$-$X$~(\AA)}& &  \\ 
\cline{3-5}\cline{7-9}\cline{11-13}\cline{15-17}\cline{19-21}
~&~&PBE&HSE&EXP&&PBE&HSE&EXP&&PBE&HSE&EXP&&PBE&HSE&EXP\\
\hline
BaCu$_{2}$S$_{2}$     & & 3.92 & 3.94 &   3.91\cite{Huster_ZAAC99}   & & 3.23 & 3.23 & 3.24\cite{Huster_ZAAC99}  & & 0.361 & 0.361 & 0.362\cite{Huster_ZAAC99}  & & 2.41 & 2.42 & 2.41\cite{Huster_ZAAC99} \\
BaAg$_{2}$S$_{2}$    & & 4.14 & 4.12 &   -   & & 3.29 & 3.33 &   -    & & 0.378 & 0.377 &  -    & & 2.70 & 2.70 &  -    \\
BaCu$_{2}$Se$_{2}$   & & 4.06 & 4.09 &   4.04\cite{Huster_ZAAC99}  & & 3.23 & 3.24 &   3.26\cite{Huster_ZAAC99}  & & 0.363 & 0.364 & 0.361\cite{Huster_ZAAC99} & & 2.52 & 2.54 & 2.50\cite{Huster_ZAAC99} \\
BaAg$_{2}$Se$_{2}$ & & 4.26 & 4.24 &   -   & & 3.33 & 3.36 &  - & & 0.378 & 0.377 &  - & & 2.80 & 2.79 &  -    \\
%SrCu$_{2}$S$_{2}$ && 3.798 & 3.798 & - & & 3.270 & 3.270 & - && 2.398 & 2.398 & - \\
%SrCu$_{2}$Se$_{2}$ && 3.924 & 3.924 & - && 3.338 & 3.338 & - && 2.526 & 2.526\\
\hline
\end{tabular}
\end{center}
\end{table*}
%The values are from ICSD

\begin{figure}
\centering
\subfloat{\includegraphics[height=2.5in]{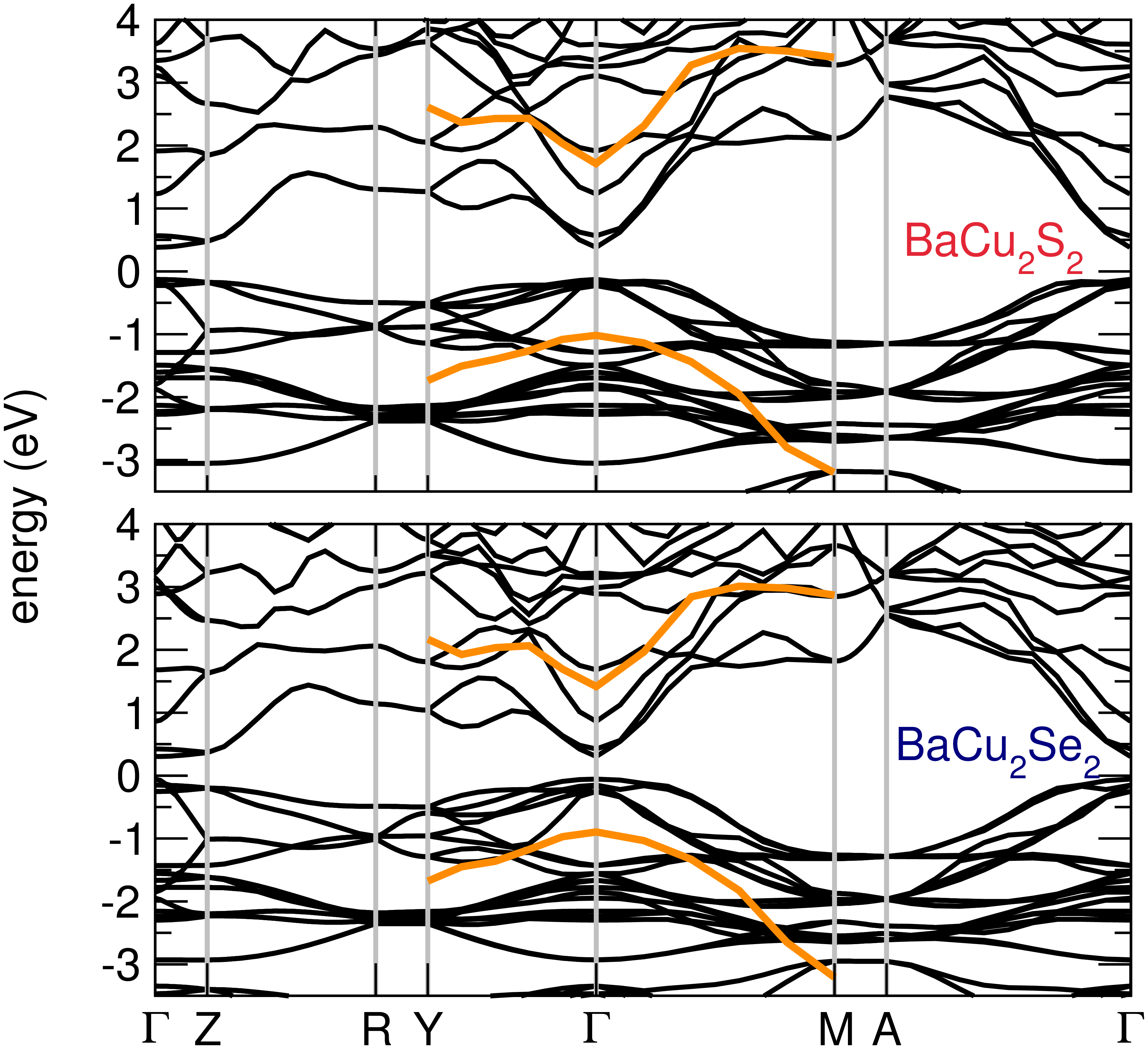}}
\subfloat{\includegraphics[height=2.5in]{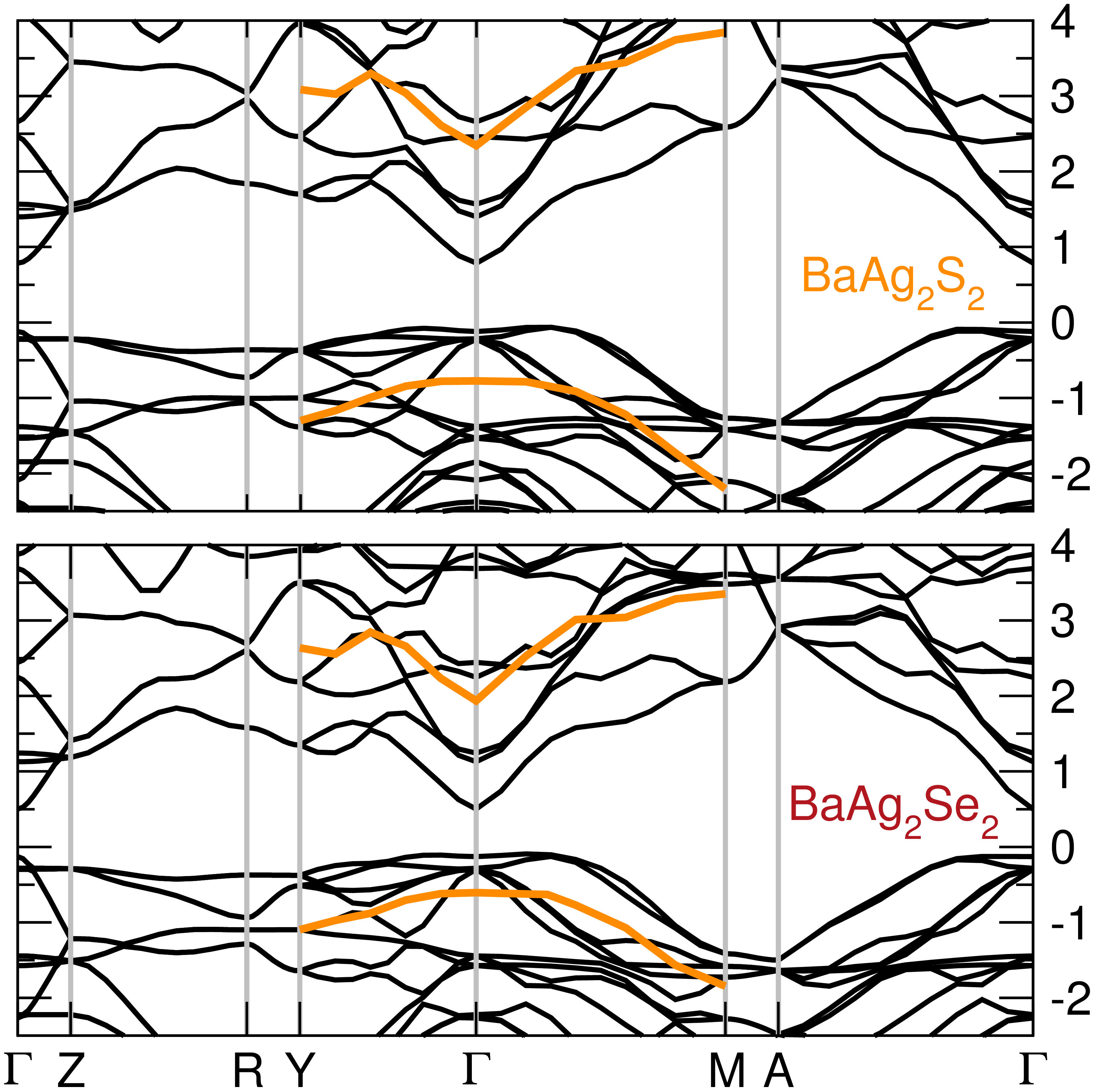}}
\caption{Electronic band structures of \bmx\/ ($M$ = Cu, Ag; $X$ = S, Se) as determined by PBE calculations, with bands from the HSE functional overlaid in bold orange lines along the Y to $\Gamma$ to M dispersion. The direct band gap occurs at the $\Gamma$ point.}
\label{bands}
\end{figure}

The calculated electronic band structures of the four compounds are shown in Fig.~\ref{bands} using both the PBE and HSE functionals, with PBE covering the full Brillouin zone and HSE only the Y to $\Gamma$ to M directions. The electronic band gap values are presented in Table~\ref{gaps}. All four compounds are direct gap semiconductors, and their valence band maxima (VBM) and their conduction band minima (CBM) occur at the $\Gamma$ point. As expected, the PBE band gaps are significantly smaller than the HSE gaps. The underestimation of the band gap is an inherent feature of local and semi-local functionals because they do not take into account the discontinuity in the exchange-correlation potential. The correction of the HSE functional to the bands happens to both conduction and valence bands (Fig.~\ref{bands}). For the Cu compounds, the downward shift of the valence bands and the upward shift of the conduction bands are similar; whereas the shift of the conduction band is more significant for Ag compounds. Replacing S with Se reduces the band gap by about 0.3\,eV and seems to make the bands more disperse; whereas replacing Cu with Ag increases the gap by about 0.4\,eV. 

\begin{table}[t]\footnotesize
%\ra{1.3}
\begin{center}
\caption{The calculated and experimental band gaps for the \bmx\/ compounds.}
\label{gaps} 
\vspace{4pt}
\begin{tabular}{c@{\hskip 3mm}ccc}
\hline
~& & $E_{\rm{g}}$ (eV) &   \\ 
\cline{2-4}
~&PBE&HSE&EXP\\
\hline
%\ra{1.3}
BaCu$_{2}$S$_{2}$ & 0.51 & 1.62 & 1.75\cite{Park_APL02}   \\ %2.3 for alpha-BaCu2S2, and 1.75 for Beta-BaCu2S2... the latter value seems to be in greater agreement with the HSE band gap obtained
BaAg$_{2}$S$_{2}$  & 0.91 & 2.01 &   -   \\
BaCu$_{2}$Se$_{2}$ & 0.36 & 1.33 &   -  \\ %Only a direct band gap for orthorhombic BaCu2Se2, not our i4/mmm structure type
BaAg$_{2}$Se$_{2}$ & 0.68 & 1.61 &  -    \\
%SrCu$_{2}$S$_{2}$  && 0.51 & 2.27 & - \\
%SrCu$_{2}$Se$_{2}$ && 0.26 & 2.03 & - \\
\hline
\end{tabular}
\end{center}
\end{table}

The experimental band gap of BaCu$_{2}$S$_{2}$ was determined to be 1.75\,eV by experiment.\cite{Park_APL02} The HSE result of 1.62\,eV matches well with this experimental value, and the slight discrepancy is likely caused by the low density of states (DOS) close to the CBM. The band gaps of the other compounds are not known experimentally. The conduction bands of the four compounds are largely dispersed around the $\Gamma$ point and are low in energy compared to conduction states in the other areas of the Brillouin zone, resulting in a very low DOS close to the CBM. This feature also occurs in many nitride and oxide semiconductors, such as InN and ZnO. For example, the gap of InN was originally measured to be 2.0\,eV for films deposited by sputtering and MOVPE,\cite{Tansley_JAP86,Guo_JJAP94} however further investigation of MBE films established the true gap of InN as 0.7 to 0.8\,eV\cite{Wu_APL02}. The small direct band gap for some of these compounds suggests that they are not ideal candidates for $p$-type transparent conducting oxides.  

\begin{figure}
\centering
\subfloat{\includegraphics[height=2.5in]{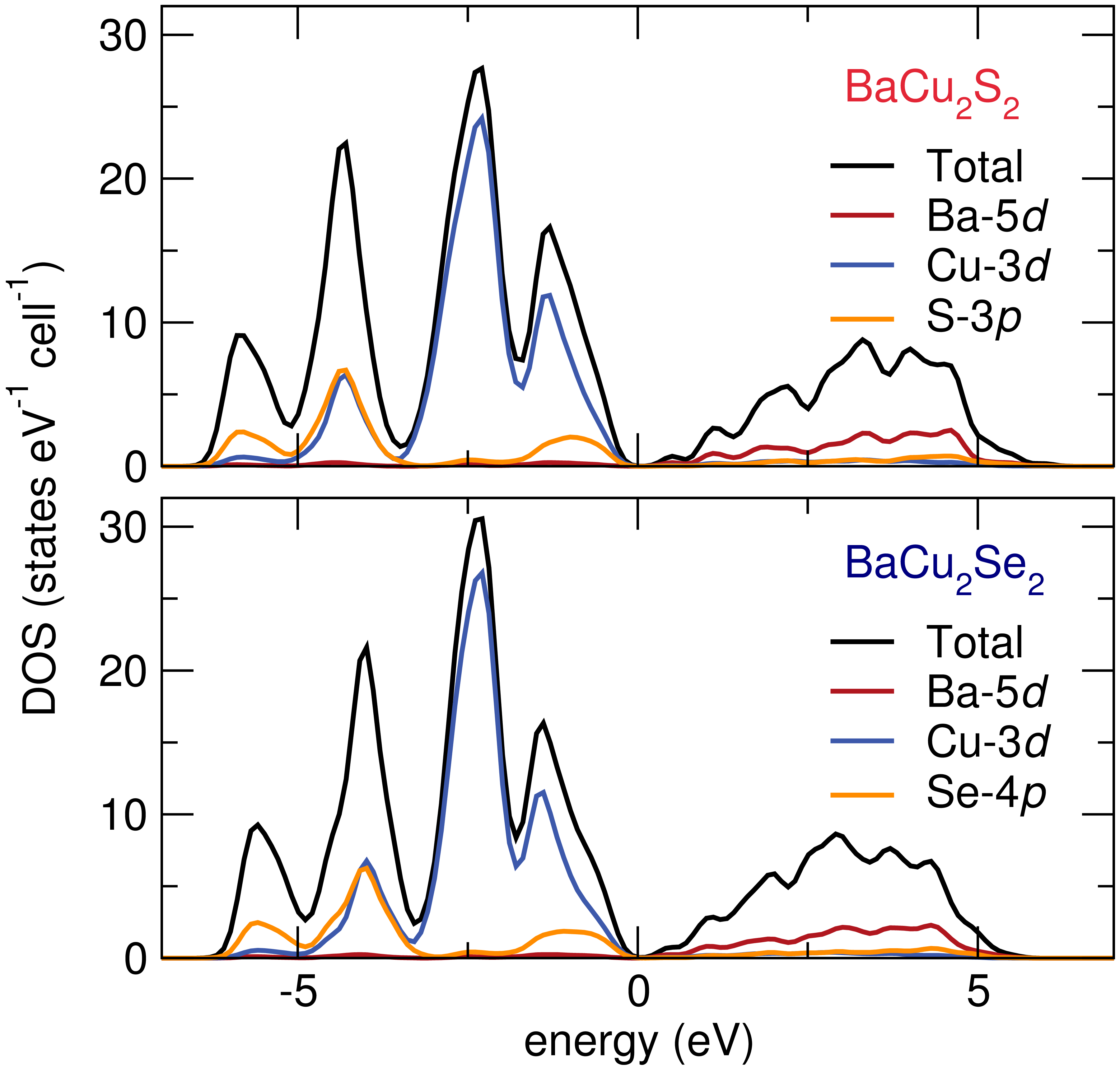}}
\subfloat{\includegraphics[height=2.5in]{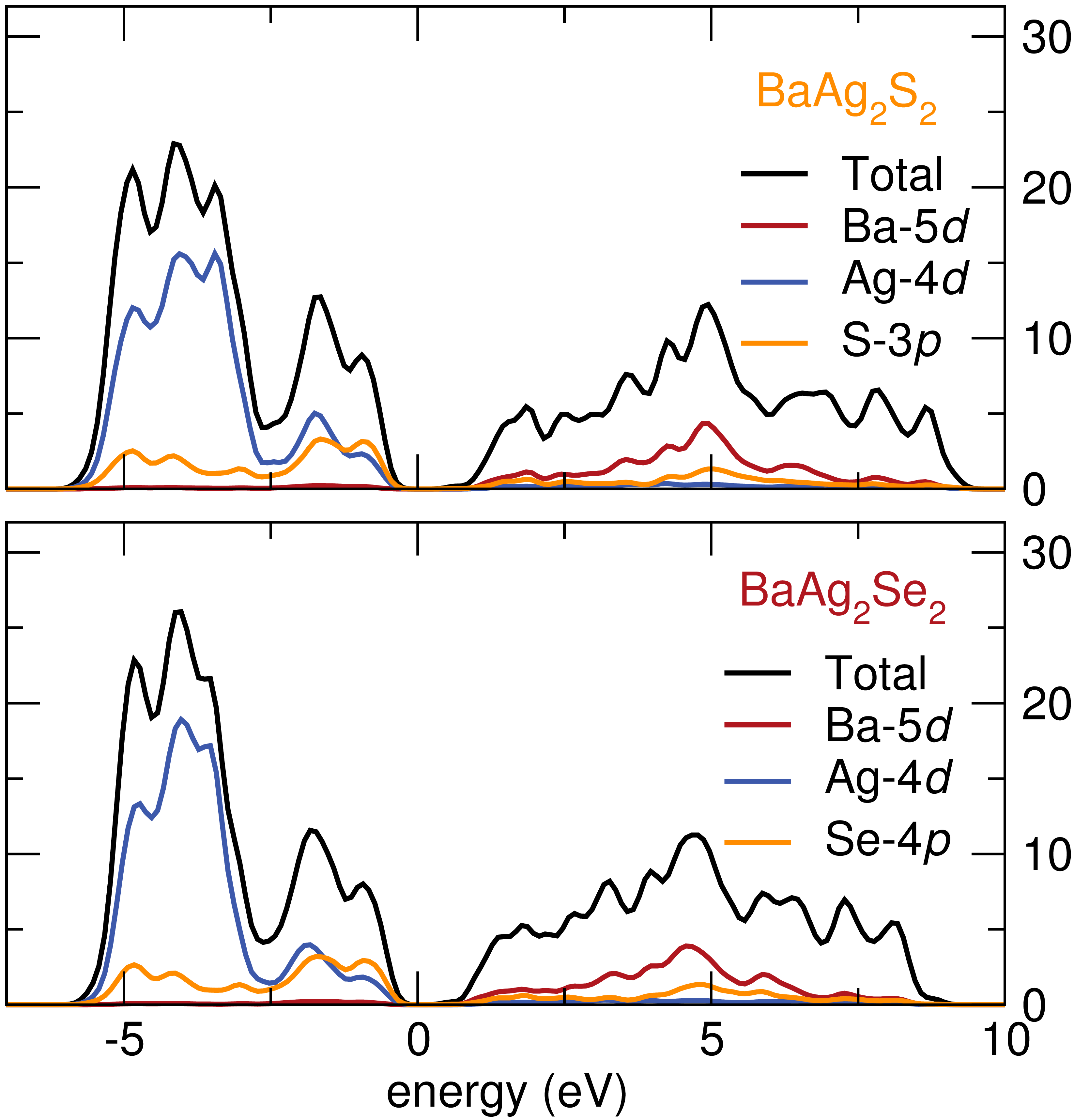}}
\caption{Electronic densities of states for \bmx\/ ($M$ = Cu, Ag; $X$ = S, Se) as determined by PBE calculations. The projected DOS are displayed for the atomic orbitals contributing near the band edges.}
\label{dos}
\end{figure}

In order to investigate the nature of the electronic states around the VBM and CBM, the wavefunctions were projected into the atomic orbitals inside the PAW radii for each element. As shown in Fig.~\ref{dos}, the valence states consist of mainly the transition metal (Cu or Ag) $d$ orbitals and the chalcogen (S or Se) $p$ orbitals. The PDOS of the sulfides and selenides with the same cations are quite similar except that in the selenides the gaps are slightly smaller and the bands are slightly broader. On the other hand, the PDOS shows very different features for the Cu and Ag compounds. In BaCu$_{2}$S$_{2}$ and BaCu$_{2}$Se$_{2}$, the states close to the VBM are dominantly Cu-$d$ states. The S-$p$ and Se-$p$ orbitals are lower in energy, at about 3 to 6\,eV below the Fermi level. In contrast, the states close to the VBM of BaAg$_{2}$S$_{2}$ and BaAg$_{2}$Se$_{2}$ consist of both Ag-$d$ and S-$p$ (Se-$p$) orbitals. The peak of the Ag-$d$ states are 3 to 5\,eV below the Fermi level. The hybridization of the $d$ and $p$ orbitals reduces the localization of hole states around the VBM and thereby increases the hole mobility.\cite{Liu_PBCM10} The states close to the CBM are composed of, to some extent unexpectedly, mainly Ba-$d$. In the Ba atom, the $s$ state is filled before $d$ in accordance with the Aufbau principle. However, the Ba-$s$ orbital has a very large radius. Upon forming the extended solid, the narrow space may push up the energy of these orbitals and become higher than that of the $d$ orbitals. A recent DFT study of the BaS electronic structure also shows that the Ba-$d$ orbitals are the dominant components of the states around the CBM.\cite{Feng_CEJ2010}

%The PBE density of states show that the bottom of the conduction band is composed mainly of the Barium-5$d$ orbital. This is in agreement with the density of states of BaS obtained by Feng et al. which also show that the Barium-5$d$ orbital is the predominant component of the conduction band \cite{Feng_CEJ2010}. This also compares well with SrCu$_2$S$_2$ and SrCu$_2$Se$_2$, where the Strontium-4$d$ orbital is the main contributor to the bottom of the conduction band. The upper valence band is composed mainly of the hybridized Cu-3$d$ and Ag-4$d$ orbitals and the S-3$p$ and Se-4$p$ orbitals. The Cu-3$d$ and Ag-4$d$ orbitals are also higher in energy than the S-3$p$ and Se-4$p$ orbitals. The hybridization of the $d$ and $p$ orbitals reduces hole localization at the edge of the valence band and thereby increases hole mobility \cite{Liu_PBCM10}. The high dispersion of the bottom of the valence band, located between -8 \,eV and 0 \,eV, and the top of the conduction band also suggest that these compounds have high electrical conductivity.

%To investigate the doping propensity of the compounds, slab model calculations were performed to find the alignment of the valence band maximum and conduction band minimum relative to the vacuum region. Since the HSE functional only makes a minimal change to the ground state distribution, it can be assumed that the energy difference between the average electrostatic potential of the bulk and vacuum levels are the same for both the PBE and HSE calculations. 

\begin{figure}
\centering
\includegraphics[width=3in]{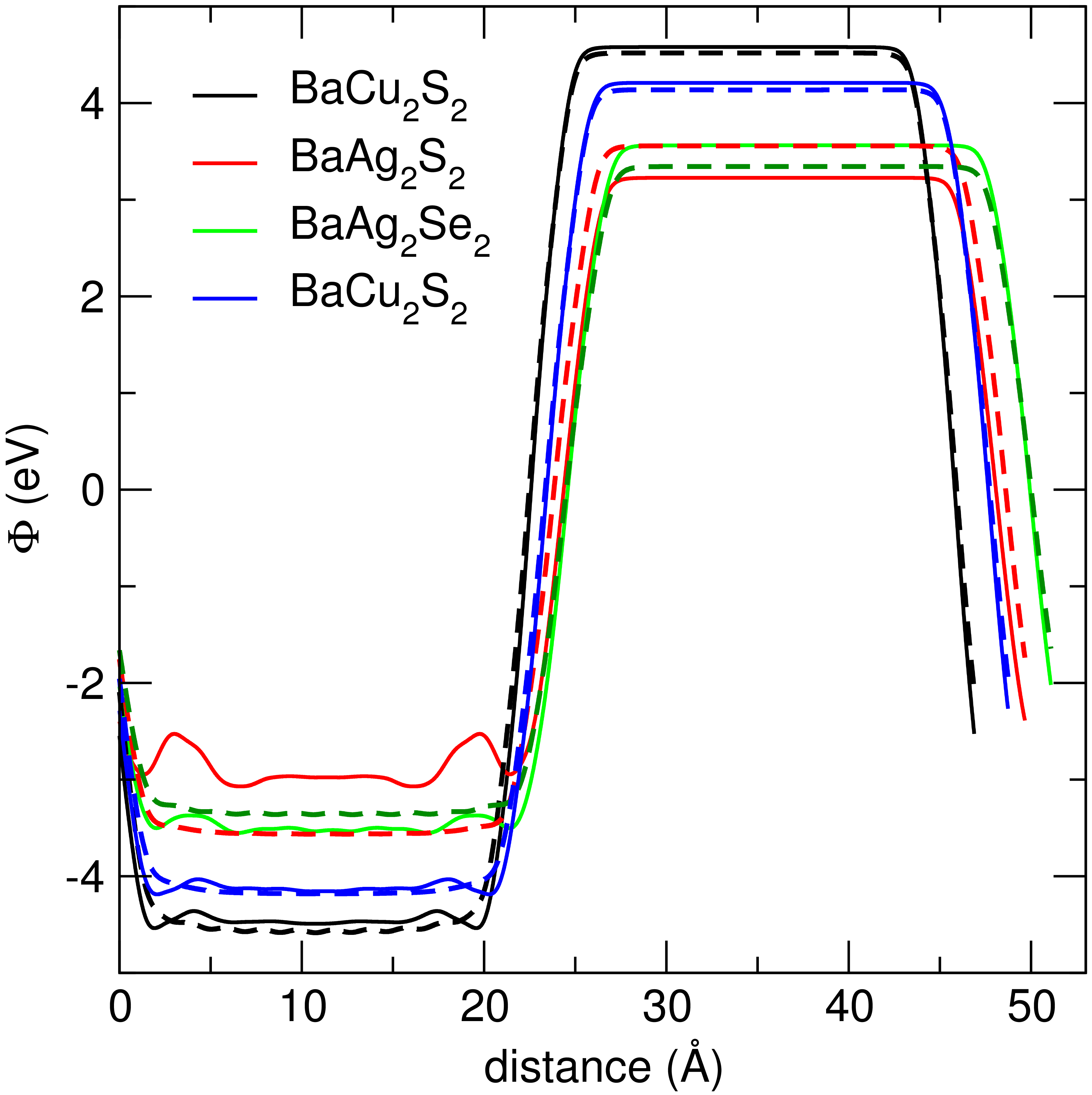}
\caption{The calculated electrostatic potential profiles for \bmx\/ ($M$ = Cu, Ag; $X$ = S, Se) as determined by slab PBE calculations. The solid and the dashed lines respectively present the results obtained by fully relaxing the atomic positions and by fixing them perpendicular to the slab, along the $c$ direction.}
\label{potential}
\end{figure}

Absolute energy alignment was achieved by aligning the average electrostatic potential of the bulk materials with that of the vacuum level by conducting slab model calculations. For these compounds, it was necessary to construct the slab model along a non-polar orientation to avoid a large polarization that can originate from the surface discontinuity. The calculated electrostatic potential profiles are shown in Fig.~\ref{potential}. The relaxation of the atoms in the c direction, perpendicular to the slab, may cause noticeable surface dipoles, causing the dips and bumps in the potential profile around the surface regions. For most of the compounds, the effect is small enough that it does not cause significant change in the potential alignment. However, the atomic relaxation at the surface is quite large for BaAg$_{2}$S$_{2}$ (red line in Fig.~\ref{potential}), causing a non-negligible change in the alignment. The band alignment difference with relaxation and non-relaxation along the perpendicular direction is as large as 0.8\,eV.  

\begin{figure}
\centering
\includegraphics[width=3in]{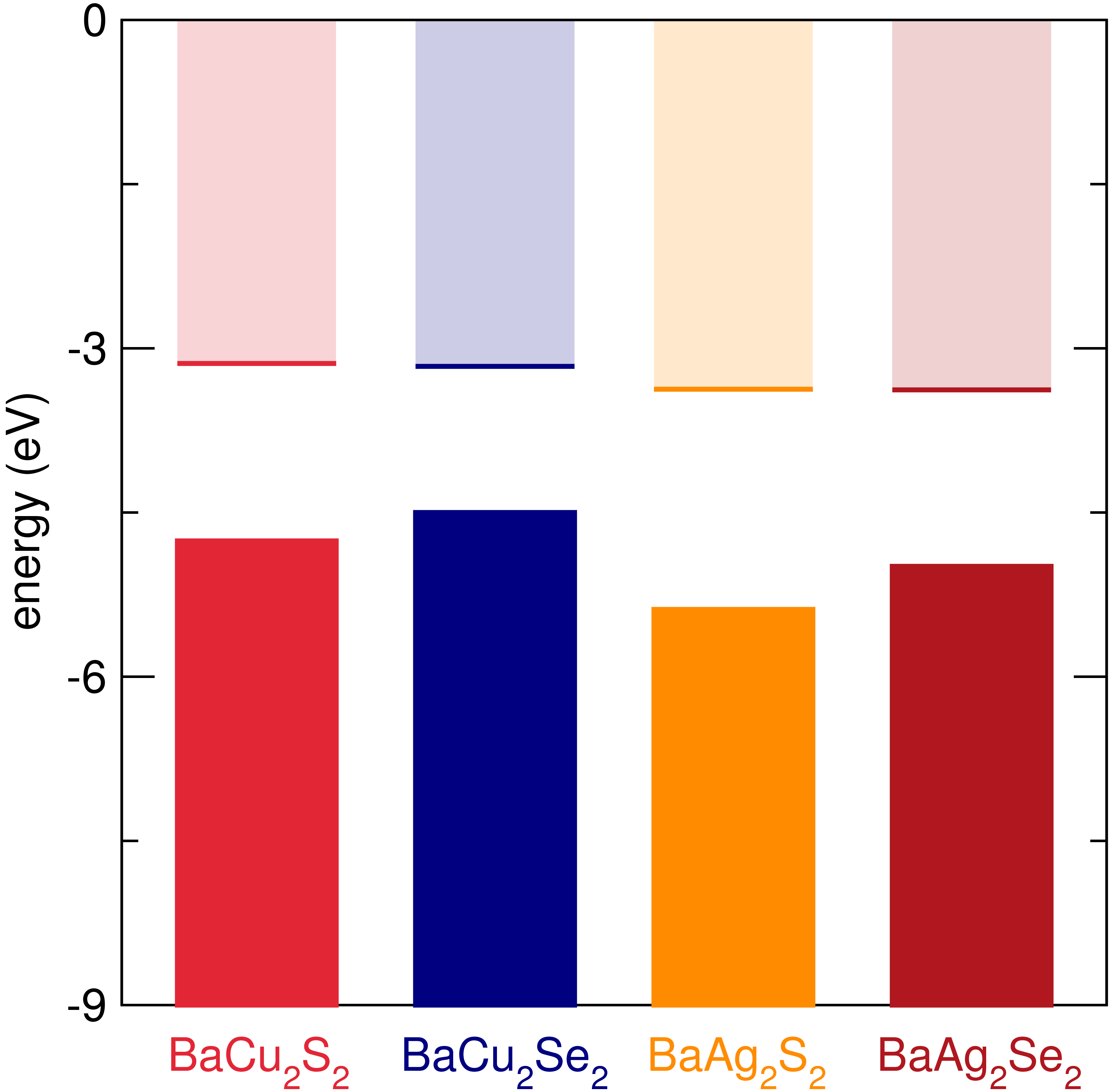}
\caption{Electronic band offset diagram for \bmx\/ ($M$ = Cu, Ag; $X$ = S, Se) as determined by slab PBE calculations. The band gaps are taken from bulk HSE calculations.}
\label{offsets}
\end{figure}

The alignments of the VBM and the CBM to the vacuum level are shown in Fig.~\ref{offsets}, and the corresponding IE and EA values are presented in Table~\ref{IE-EA}. In comparison with oxides and nitrides, for which the $p$-type doping processes are impeded by exceedingly large IEs, the IE and EA values are quite small for these four sulfide and selenide compounds. Their IE values vary over a small energy range from 5\,eV to 4.5\,eV. GaN has an IE value of about 6.5\,eV, and is well known for its difficulty to dope $p$-type\cite{Calleja_JCG1999}. The success in overcoming this issue by doping Mg in GaN was the major step towards the fabrication of blue and UV nitride LED and laser diodes\cite{Strite_JVSTB1992}. The IE is even higher for ZnO (about 8\,eV)\cite{Limpijumnong_PRL04} and its $p$-type doping is still the major challenge for its application as an optoelectronic material. On the other hand, it may be difficult to $n$-type dope the \bmx\/ materials of this study because their EA values are very low. Indeed, since the states around the CBM mainly consist of Ba-$d$ states, one needs to transfer the electrons from donor states into these Ba-$d$ states in order to make $n$-type doping possible.

The band alignments are quite similar for the four compounds. The variation of the VBM position is less than 0.5\,eV. In general, the Ag-$d$ levels are lower than the Cu-$d$ levels. However, comparing the PDOS of the compounds with the different transition metals (Fig. \ref{dos}), the Cu-$d$ states dominate the edge of the VBM, whereas the Ag-$d$ states are several eV lower than the VBM. The same difference between Cu and Ag occurs in the related layered delafossite oxides.\cite{Kandpal_SSS02}

\begin{table}[t]\footnotesize
%\ra{1.3}
\begin{center}
\caption{The calculated ionization energies and electron affinities for the \bmx\/ compounds. Experimental values are unavailable.}
\label{IE-EA} 
\vspace{4pt}
\begin{tabular}{p{0.7cm}p{0.7cm}p{0.7cm}p{0.7cm}p{0.7cm}p{0.7cm}}
\hline
~&~&$E_{\rm{I}}$~(eV) && $E_{\rm{A}}$~(eV) \\ 
\cline{3-4}\cline{5-6}
~&~&PBE&HSE&PBE&HSE\\
\hline
%\ra{1.3}
BaCu$_{2}$S$_{2}$   & &3.78 & 4.76 & 3.27 & 3.14 \\
BaAg$_{2}$S$_{2}$   & &4.71 & 5.39 & 3.80 & 3.37   \\
BaCu$_{2}$Se$_{2}$ & &3.67 & 4.50 & 3.31 & 3.17  \\
BaAg$_{2}$Se$_{2}$ & &4.43 & 4.99 & 3.75 & 3.38  \\
\hline
\end{tabular}
\end{center}
\end{table}

\section{Conclusion}

Using the DFT method, with both semi-local and hybrid functionals, the electronic structures and band alignments of \bcu\/ and \bce\/, and the theoretical structures of \bau\/ and \bae\/, were calculated, including their ionization energies and electron affinities. The results show that the band gaps of these compounds are somewhat low, between 1.33 and 2.01\,eV, suggesting they might not be good candidates as transparent conducting materials. The low ionization energies and the small electron affinities of these compounds suggest a strong propensity of $p$-type doping in these compounds. While the replacement of S by Se decreases the band gap, the replacement of Cu by Ag significantly increases the band gap and keeps a similar ionization energy, therefore retaining the tendency toward $p$-type doping.

\section{Acknowledgements}

This project was supported by the NSF through DMR 1105301. PTB is supported by the NSF Graduate Research Fellowship Program. MSM is supported by the ConvEne-IGERT Program (NSF-DGE0801627). AK is supported by the Materials Research Laboratory RISE Scholarship. The work made use of the computing facilities of the Center for Scientific Computing supported by the California Nanosystems Institute (Grant CNS-0960316), Hewlett-Packard, and by the Materials Research Laboratory at UCSB: an NSF MRSEC (Grant DMR-1121053). Part of the calculations also made use of computing facilities supported by the NSF-funded XSEDE under Grant TG-DMR130005.

\clearpage

\end{document}